\documentclass{llncs}
\usepackage{graphicx}
\usepackage{amsmath,amssymb} 
\newcommand{\ket}[1]{ \left |#1\right \rangle}
\makeindex

\begin{document}

\setcounter{page}{1}
\pagestyle{headings}  

\title{State Transfer instead of Teleportation in Measurement-based Quantum Computation}
\author{Simon Perdrix}
\institute{Leibniz Laboratory\\46 avenue F\'elix Viallet 38000 Grenoble, France\\\emph{simon.perdrix@imag.fr}}
\maketitle
\begin{abstract}
Quantum measurement is universal for quantum computation (Nielsen \cite{N01}, Leung \cite{L01,L03}). The model of quantum computation introduced by Nielsen and further developed by Leung relies on a generalized form of teleportation. In order to simulate any $n$-qubit unitary transformation with this model, $4$ auxiliary qubits are required. Moreover Leung exhibited a universal family of observables composed of $4$ two-qubit measurements. We introduce a model of quantum computation via measurements only, relying on \emph{state transfer}: state transfer only retains the part of teleportation which is necessary for computating.  In order to simulate any $n$-qubit unitary transformation with this new model, only one auxiliary qubit is required. Moreover we exhibit a universal family of observables composed of $3$ one-qubit measurements and only one two-qubit measurement. This model improves those of Nielsen and Leung in terms of both the number of auxi\-liary qubits and the number of two-qubit measurements required for quantum universality. In both cases, the minimal amounts of necessary resources are now reached: one auxiliary qubit (because measurement is destructive) and one two-qubit measurement (for creating entanglement).

\end{abstract}

\section{Quantum computation via measurements only based on teleportation}

The proof of the universality of measurement-based quantum computation proceeds by simulating all quantum circuits. This proof is based on a 3-level decomposition of the problem: \emph{steps of simulation of U}, where a unitary transformation $U$ is simulated up to a Pauli operator (i.e. a step of simulation of $U$ on $\ket{\phi}$ produces $\sigma U \ket{\phi}$, where $\sigma$ is a Pauli operator); \emph{full simulation of U}, which combines a step of simulation of $U$ with steps of simulation of Pauli operators; and \emph{simulation of a quantum circuit}, which combines full simulations of unitary transformations.

\subsection{From teleportation with measurements only to a step of simulation of a unitary transformation}

\begin{center}
\includegraphics[width=0.5\textwidth]{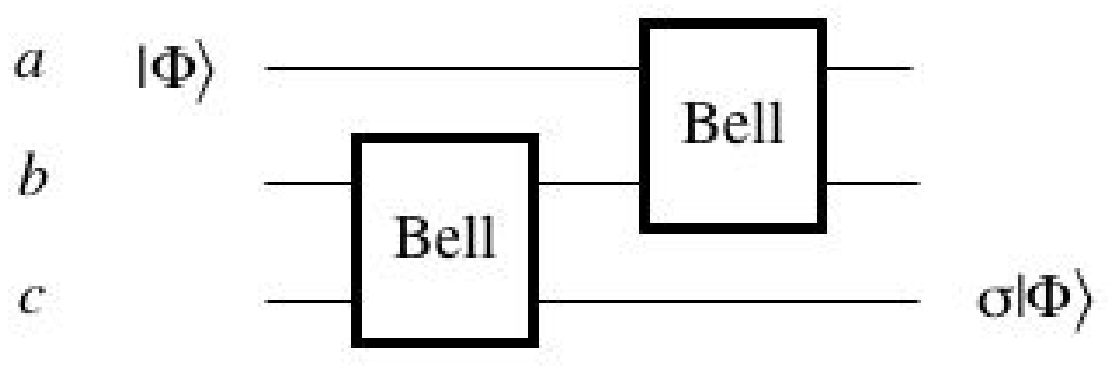}

\emph{\small Figure 1 - Teleportation with measurements only}
\end{center}

The state $\ket{\phi}$ of a qubit $a$ can be teleported to a qubit $c$ using Bell measurements only (fig. 1) and an additional qubit $b$. A Bell state is created with a Bell measurement on the two auxiliary qubits $b$ and $c$, then a Bell measurement is performed on qubits $a$ and $b$ for teleporting $\ket{\phi}$ from $a$ to $c$. For a given 1-qubit unitary operator $U$, by modifying the basis of the measurement performed during teleportation (fig. 2), a \emph{step of simulation of $U$} (i.e. a simulation of $U$ up to a Pauli operator) is obtained. This step of simulation needs two auxiliary qubits. An extension to the simulation of 2-qubit unitary transformations (e.g. $CNot$) may be done using only 2-qubit measurements \cite{L01}. These simulations require four auxiliary qubits. A step of simulation of $U$ can be abstracted into a black box (fig. $3$) with one input and four outputs according to the four Pauli operators, for use at the next higher level which is the full simulation of $U$. 

\begin{center}
\includegraphics[width=0.9\textwidth]{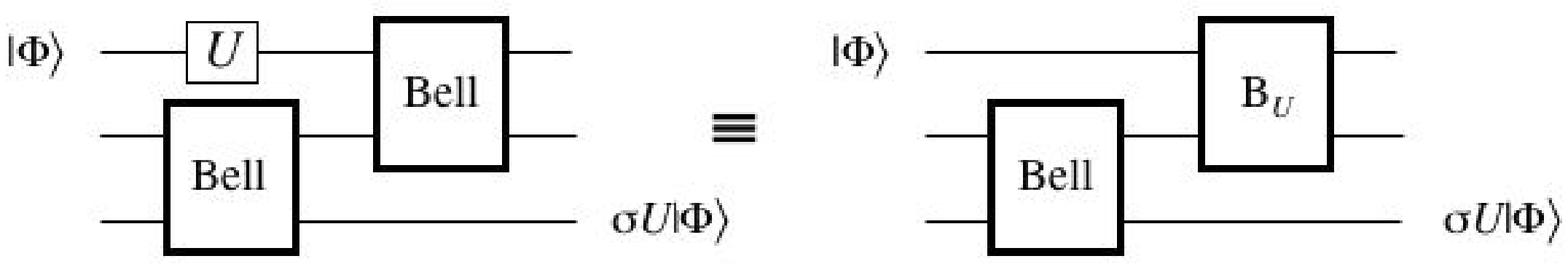}

\emph{\small Figure 2 - Left: Teleportation of $U\ket{\phi}$; Right: a step of simulation of U}
\end{center}

\begin{center}

\includegraphics[width=0.15\textwidth]{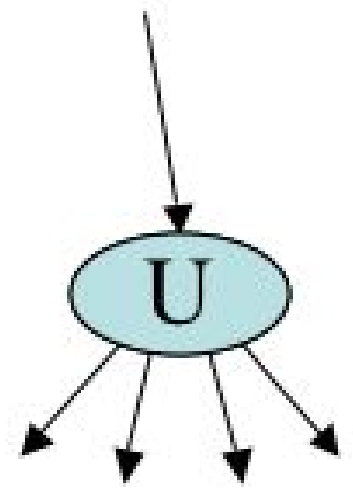}

\emph{\small Figure 3 - General black-box representation of a step of simulation of $U$.}
\end{center}

\subsection{Full simulation of a unitary transformation}

For a given step of simulation of $U$, the \emph{full simulation of $U$} is given by an automaton where each state encapsulates a step of simulation (fig. $4$). This automaton is interpreted as follows: $U$ is simulated on a quantum state $\ket{\phi}$ so $\sigma U\ket{\phi}$ is obtained, where $\sigma$ is a Pauli operator. If $\sigma=I$ then the simulation is terminated, otherwise $\sigma$ is simulated. From this step of simulation, the state $\sigma'\sigma \sigma U\ket{\phi}=\sigma' U\ket{\phi}$ is obtained. If $\sigma'=I$ the simulation is terminated, otherwise $\sigma'$ is simulated, and so on.

\begin{center}
\includegraphics[width=0.4\textwidth]{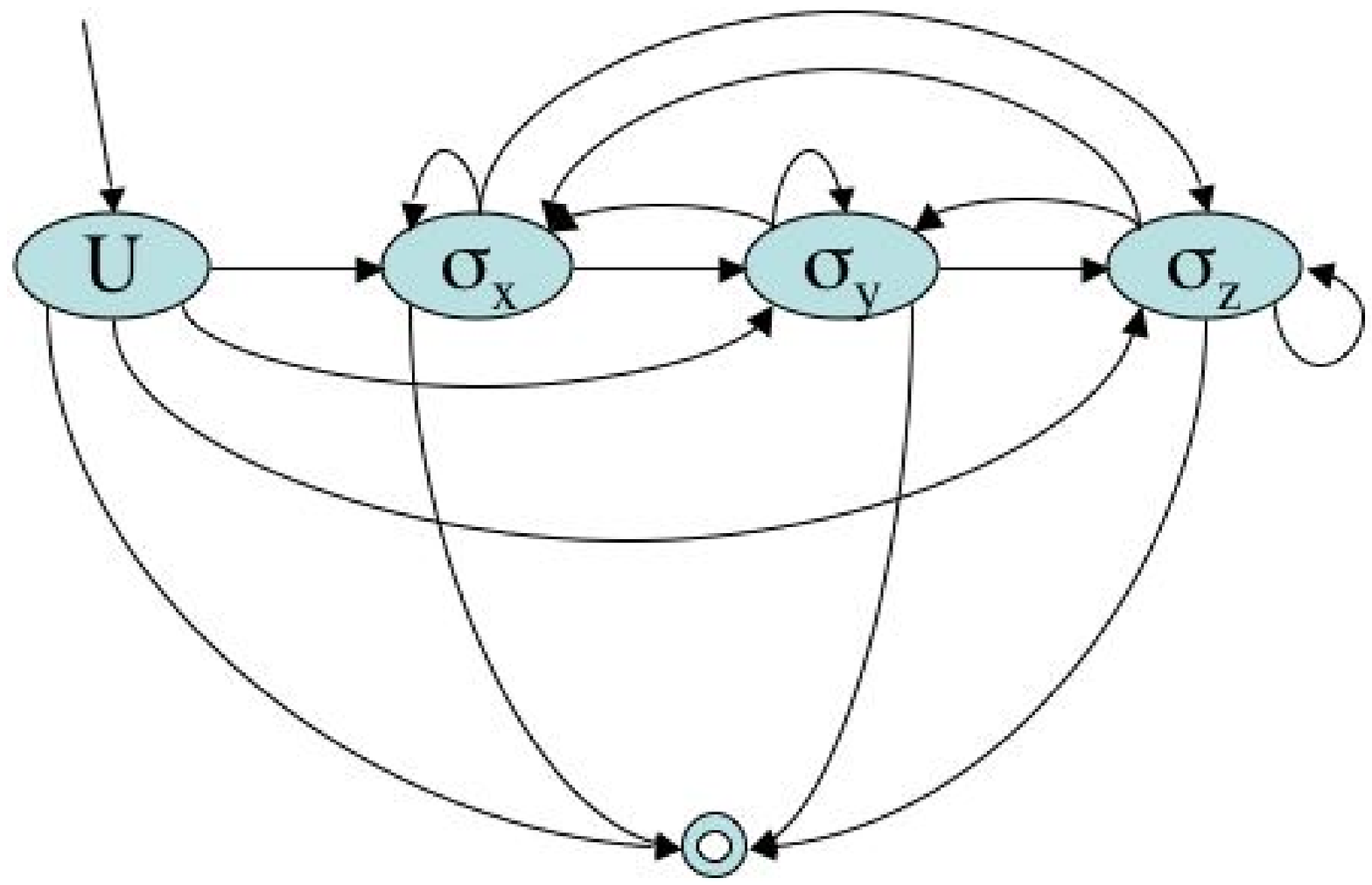}

\emph{\small Figure 4 - Full simulation of $U$}
\end{center}

\subsection{Universality}

The model of quantum circuits based on the family of unitary transformations $\{H,T,CNot\}$ is quantum universal (see \cite{KSV} for details on quantum circuits and \cite{PJ04} for details on the notions of universalities and simulations). Thus, since all unitary transformations can be full simulated with measurements only, the model composed of automata where the states are steps of simulation is quantum universal.

\begin{center}
$H=\frac{1}{\sqrt{2}}\left(\begin{array}{cc}
  1 & 1\\
  1 & -1\\
\end{array} \right)$, 
$T=\left(\begin{array}{cc}
  1 & 0\\
  0&e^{\frac{i\pi}{4}}\\
\end{array} \right)$, 
$CNot=\left(\begin{array}{cccc}
  1 & 0&0&0\\
0 & 1&0&0\\
0 & 0&0&1\\
0& 0&1&0
\end{array} \right)$
\end{center}

\subsection{Resources}
For each step of simulation of a $1$-qubit (resp $2$-qubit) unitary transformation, two (resp four) auxiliary qubits are needed. 

In terms of observables, only $2$-qubit projective measurements are used in the scheme developed by Leung. Moreover the minimal set of observables needed for quantum universality with this scheme comprises four elements: $\{X\otimes X, Z \otimes Z, X\otimes Z, \frac{1}{\sqrt{2}}(X+Y)\otimes X\}$\cite{L01}.

\section{Quantum computation via measurements only based on state transfer}
Whereas the scheme of computation by Nielsen and improved by Leung is based on teleportation, we introduce a measurement-based quantum computation based on \emph{state transfer}, which is an alternative to teleportation for purpose of computation. State transfer needs less measurements and less auxiliary qubits than teleportation, but on the other hand, state transfer cannot replace teleportation in \emph{non-local} applications. 

\subsection{State transfer}

\begin{center}
\includegraphics[width=0.5\textwidth]{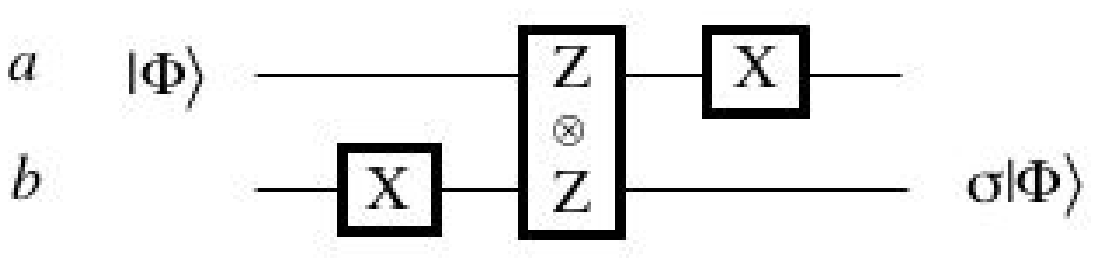}

\emph{\small Figure 5 - State transfer}
\end{center}

\begin{lemma}
For a given qubit $a$ and an auxiliary qubit $b$, the sequence of measurements $\{X^{(b)}, Z^{(a)}\otimes Z^{(b)}, X^{(a)}\}$ (see fig. 5), transfers the state $\ket{\phi}=\alpha\ket{0}+\beta\ket{1}$ from $a$ to $b$ up to a Pauli operator which depends on the classical outcomes of the measurements.
\end{lemma}

\begin{proof}

Since a $X^{(b)}$-measurement is a 1-qubit measurement of $b$ in the basis given by eigenvalues of $X$, i.e. $\{ \frac{\ket{0}+\ket{1}}{\sqrt{2}},\frac{\ket{0}-\ket{1}}{\sqrt{2}}\}$, the state $\ket{\psi_1}$ of the register $a,b$ after this measurement is:

 $\ket{\psi_1}=(\alpha \ket{0}+\beta \ket{1})\otimes(\sigma_z^\frac{1-j}{2}(\frac{\ket{0}+\ket{1}}{\sqrt{2}}))$,
 
 $\ket{\psi_1}=\frac{1}{\sqrt{2}}(I\otimes \sigma_z^\frac{1-j}{2}) [\alpha \ket{00}+\alpha \ket{01}+\beta \ket{10}+\beta \ket{11}]$,

where $j\in\{-1,1\}$ is the outcome of the $X^{(b)}$-measurement. 
A $Z^{(a)}\otimes Z^{(b)}$-measurement is a projective two-qubit measurement with two possible classical outcomes $1$ and $-1$. If this outcome is $k$, then the state $\ket{\psi_2}$ of the register $a,b$ after this measurement is:

 $\ket{\psi_2}=(\sigma_z^\frac{1-j}{2} \otimes \sigma_x^\frac{1-k}{2}) [\alpha \ket{00}+\beta \ket{11}]$.
 
 If the outcome of the $X^{(a)}$-measurement is $l\in \{-1,1\}$, then the state $\ket{\psi_3}$ of the register $a,b$ after this measurement is:
 
 $\ket{\psi_3}=\frac{1}{\sqrt{2}}(\sigma_z^\frac{1-l}{2}\otimes \sigma_z^\frac{1-j.l}{2}\sigma_x^\frac{1-k}{2}) [\frac{\ket{0}+\ket{1}}{\sqrt{2}}\otimes (\alpha \ket{0}+\beta \ket{1})]$.

Thus the state $\ket{\phi}=\alpha\ket{0}+\beta\ket{1}$ is transfered from $a$ to $b$ up to a Pauli operator $\sigma=\sigma_z^\frac{1-j.l}{2}\sigma_x^\frac{1-k}{2}$.$\hfill \Box$

 \end{proof}
 
 \subsection{Step of simulation based on State Tranfers}

By modifying the measurements performed during state transfer, all $1$-qubit unitary transformations $U$ may be simulated up to a Pauli operator using generalized state transfers. For use in later developments, a general scheme with two unitary transformations $U$ and $V$ is introduced, see fig. 6 and fig. 7.

 \begin{center}
\includegraphics[width=0.5\textwidth]{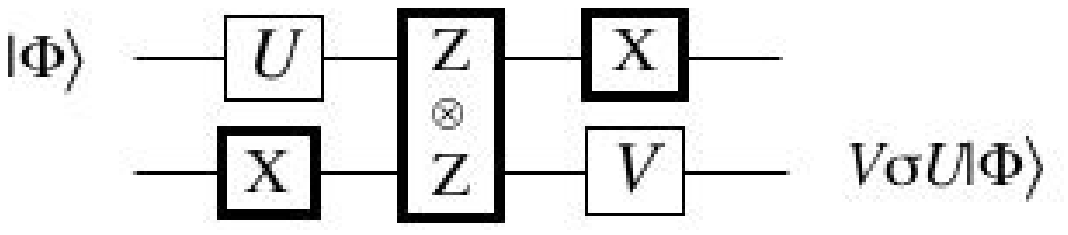}

\emph{\small Figure 6 - State Transfer with additional unitary transformations $U$ and $V$.}
$$ $$
\newpage
\includegraphics[width=0.7\textwidth]{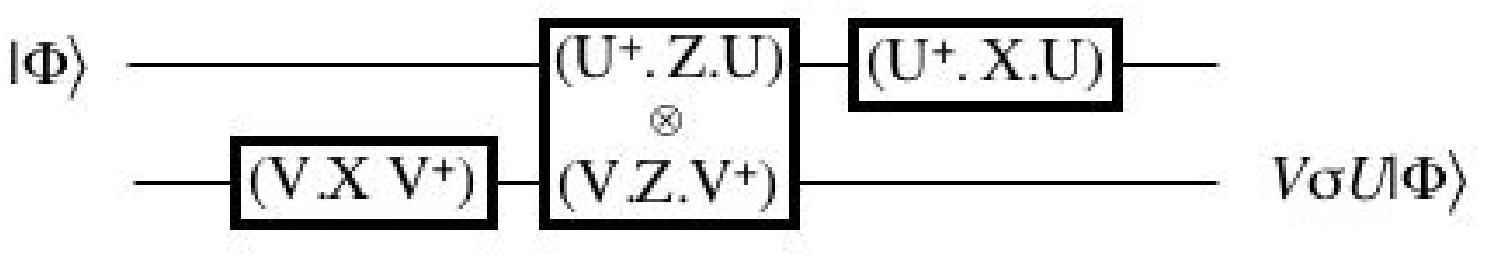}

\emph{\small Figure 7 - Generalized State Transfer.}
\end{center}

For instance, the generalized state transfers which simulate $H$, $T$ and $H.T$ are given in fig. 8, 9 and 10.

\begin{center}
\includegraphics[width=0.9\textwidth]{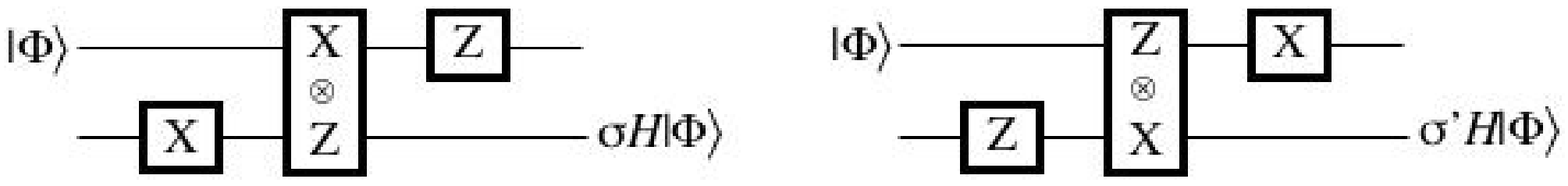}

\emph{\small Figure 8 - Steps of simulation of H - Left: $U=H$ and $V=Id$ - Right: $U=Id$ and $V=H$ (note that for all $\sigma$, there exists $\sigma '$ such that $H\sigma=\sigma ' H$)}

$$ $$

\includegraphics[width=0.6\textwidth]{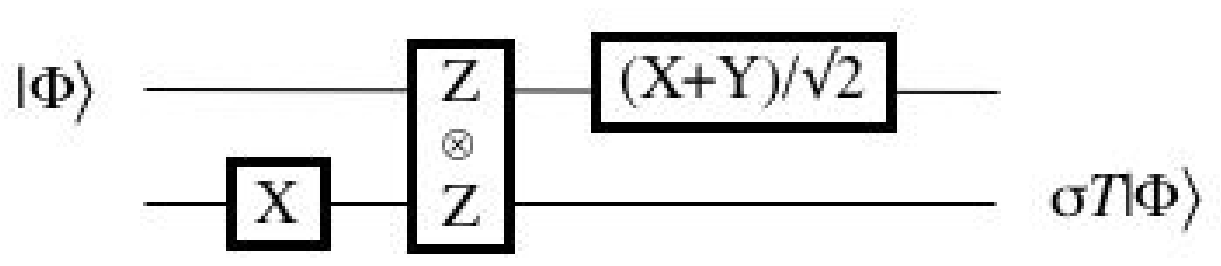}

\emph{\small Figure 9 - Step of simulation of $T$: $U=T$ and $V=Id$.}

$$ $$

\includegraphics[width=0.6\textwidth]{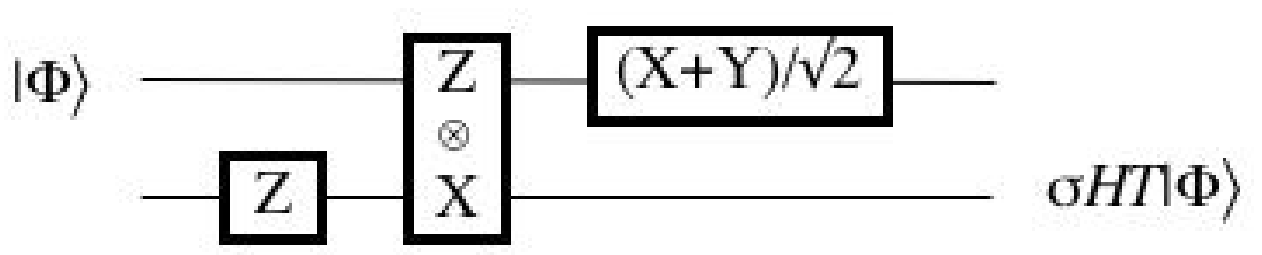}

\emph{\small Figure 10 - Step of simulation of $HT$: $U=T$ and $V=H$.}

\end{center}

\begin{lemma}
 For a given $2$-qubit register $a,b$ and one auxiliary qubit $c$, the sequence of measurements $\{Z^{(c)},Z^{(a)}\otimes X^{(c)}, Z^{(c)}\otimes X^{(b)}, X^{(c)}\}$ (see fig. 11), simulates the $2$-qubit unitary transformation $CNot$ on the state $\ket{\phi}$ of $a,b$ up to a $2$-qubit Pauli operator which depends on the classical outcomes of the measurements.
\end{lemma}
  
\begin{center}
\includegraphics[width=0.8\textwidth]{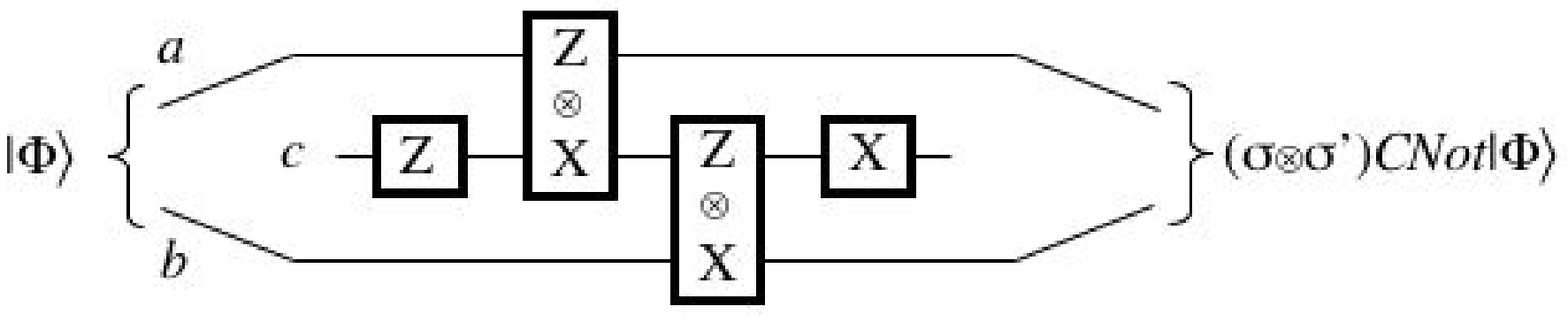}

\emph{\small Figure 11 - Step of simulation of $CNot$}
\end{center}

\begin{proof}
 
The notations $\ket{+}=\frac{\ket{0}+\ket{1}}{\sqrt{2}}$ and $\ket{-}=\frac{\ket{0}-\ket{1}}{\sqrt{2}}$ are used in this proof.

If the state $\ket{\phi}$ of the register $a,b$ is $\alpha\ket{00}+\beta\ket{01}+\gamma \ket{10} +\delta\ket{11}$, then the state $\ket{\psi_1}$ of the register $a,b,c$ after the $Z^{(c)}$-measurement is:

$\ket{\psi_1}=(I\otimes I \otimes \sigma_x^\frac{1-j}{2})[(\alpha\ket{000}+\beta\ket{010}+\gamma \ket{100} +\delta\ket{110}]$,

where $j\in \{-1,1\}$ is the classical outcome of the $Z^{(c)}$-measurement. If the outcome of the $Z^{(a)}\otimes X^{(c)}$-measurement is $k\in \{-1,1\}$, then the state $\ket{\psi_2}$ of the register $a,b,c$ after this measurement is:

$\ket{\psi_2}=(\sigma_z^\frac{1-j}{2}\otimes I \otimes \sigma_z^\frac{1-k}{2})[\alpha\ket{00+}+\beta\ket{01+}+\gamma \ket{10-}+\delta\ket{11-}]$.

If the outcome of the $Z^{(c)}\otimes X^{(b)}$-measurement is $l\in \{-1,1\}$, then the state $\ket{\psi_3}$ of the register $a,b,c$ after this measurement is:

$\ket{\psi_3}=\frac{1}{\sqrt{2}}(\sigma_z^\frac{1-j.l}{2}\otimes \sigma_x^\frac{1-k}{2} \otimes \sigma_x^\frac{1-l}{2})[\alpha(\ket{00+}+\ket{01-})+\beta(\ket{01+}+\ket{00-})+\gamma (\ket{10-}+\ket{11+})+\delta(\ket{11-}+\ket{10+})]$.

If the outcome of the $X^{(c)}$-measurement is $m\in \{-1,1\}$, then the state $\ket{\psi_4}$ of the register $a,b,c$ after this measurement is:

$\ket{\psi_4}=(\sigma_z^\frac{1-j.l}{2}\otimes \sigma_x^\frac{1-k.m}{2} \otimes \sigma_z^\frac{1-m}{2}) [(\alpha\ket{00}+\beta\ket{01}+\gamma \ket{11}+\delta\ket{10})\otimes \ket{+}]$.

Since $CNot\ket{\phi}=\alpha\ket{00}+\beta\ket{01}+\gamma \ket{11} +\delta\ket{10}$, the $2$-qubit unitary transformation \emph{CNot} is simulated up to a $2$-qubit Pauli operator.$\hfill \Box$

\end{proof}
 
\subsection{Universality}
 
Since the family of unitary transformations $\{H,T, CNot\}$ is universal, the proof of the universality of quantum computation based on state transfer is the same as the proof given in section $1$, where steps of simulation based on teleportation are replaced by steps of simulation based on state transfer.
 
\subsection{Resources}
 
 For each step of simulation of a $1$-qubit or $2$-qubit unitary transformation, only one auxiliary qubit is needed. Since the simulation with measurements only of a given unitary transformation on an unknown state cannot be performed without auxiliary qubit, the minimal resources necessary for quantum universality in terms of auxiliary qubits are reached. 

In terms of observables, only $2$-qubit projective measurements are used in the scheme based on state transfer. Since $1$-qubit observables are not quantum universal, the minimal bound of the resources needed for quantum universality is reached.

Moreover, in order to simulate the universal family of unitary transformations $\{H,$ $T,$ $CNot\}$, the steps of simulation of $H$, $T$, $CNot$ and $Identity$ are sufficient. The step of simulation of $Identity$, which is nothing but basic state transfer (fig. 5) is used to simulate Pauli operators. Thus the family of observables $\mathcal{O}_1=\{Z,X,\frac{1}{\sqrt{2}}(X+Y), Z\otimes Z, Z\otimes X\}$ is quantum universal.

In order to obtain a smaller family of observables, the family of unitary transformations  $\{H,HT,CNot\}$, which is trivially universal, is considered. Since a step of simulation of $Identity$ can be obtained by two successive steps of simulation of $H$, the family of observables $\mathcal{O}_2=\{Z,X,\frac{1}{\sqrt{2}}(X+Y), Z\otimes X\}$ is universal.

Since $1$-qubit observables are not quantum universal, all universal families of observables contain at least one two-qubit observable. Thus, since $\mathcal{O}_2$ contains only one two-qubit observable, the minimal resources necessary for quantum universality in terms of the number of different two-qubit observables, are reached.

\section{Comparison and Conclusion}

We introduced a model of quantum computation via measurements only based on state transfer. This model improves those of Nielsen and Leung in terms of the number of auxiliary qubits necessary for quantum universality of measurement-based quantum computation, and in terms of universal families of observables. In both cases, the minimal resources are now reached.


\begin{thebibliography}{99}
\bibitem{KSV}A. Y. Kitaev, A. H. Shen and M. N. Vyalyi. {\it Classical and Quantum Computation}, American Mathematical Society, 2002. 
\bibitem{N01} M. A. Nielsen. {\it Universal quantum computation using only projective measurement, quantum memory, and preparation of the 0 state}, arXiv.org report quant-ph/0108020, 2001.
\bibitem{L01} D. W. Leung. {\it Two-qubit projective measurements are universal for quantum computation}, arXiv.org report quant-ph/0111077, 2001.
\bibitem{L03} D. W. Leung. {\it Quantum computation by measurements}, arXiv.org report quant-ph/0310189, 2003.
\bibitem{PJ04} S. Perdrix and Ph. Jorrand. {\it Measurement-based Quantum Turing Machines and Questions of  Universalities
}, arXiv.org report quant-ph/0402156, 2004.
\end{thebibliography}
\end{document}